\documentclass[aps,prb,twocolumn,showpacs,groupedaddress]{revtex4}
\usepackage{graphicx}

\begin{document}

\title{Revealing polarons with high pressure on low electron-doped manganites}

\author{G. Garbarino}
\thanks{scholarship of CONICET of Argentina}
\author{C. Acha}
\thanks {fellow of CONICET of Argentina}
\email{acha@df.uba.ar} \affiliation{Laboratorio de Bajas Temperaturas, Depto.
de F\'{\i}sica, FCEyN, Universidad de Buenos Aires, Pabell\'on I, Ciudad
Universitaria, 1428 Buenos Aires, Argentina}
\author{D. Vega}
\author{G. Leyva}
\author{G. Polla}
\affiliation{Departamento de F\'{\i}sica, Comisi\'on Nacional de Energ\'{\i}a
At\'omica, Gral Paz 1499, 1650 San Mart\'{\i}n, Buenos Aires, Argentina}
\author{C. Martin}
\author{A. Maignan}
\author{B. Raveau}
\affiliation{Laboratoire CRISMAT, UMR 6508, ISMRA, 14050 Caen Cedex, France}



\date{\today}

\draft

\begin{abstract}

Pressure sensitivity (up to 1 GPa) of the electrical resistivity and of the ac
susceptibility was measured for low electron doping levels of
Ca$_{1-x}$Y$_x$MnO$_3$ (CYMO) and of Ca$_{1-x}$Sm$_x$MnO$_3$ (CSMO) ceramic
samples (0.05 $\leq$ x $\leq$ 0.15). A very weak pressure dependence of the
Curie temperature (T$_c$) was observed for both systems ($\sim$ 6
K~GPa$^{-1}$), when compared to the hole-doped manganites of the same T$_c$
($\sim$ 20 K~GPa$^{-1}$). Our results can be interpreted within a modified
Double Exchange scenario, where pressure alters the reduction of the bandwidth
produced by the electron-phonon interaction associated with small Fr\"{o}hlich
polarons in the weak to intermediate coupling regime.

\end{abstract}

\maketitle

\section{INTRODUCTION}

The study of high pressure effects on the magnetic transitions and the
electrical transport properties of manganite compounds can provide useful
information about the relevant mechanism in magnetic ordering and its relation
to transport properties. Indeed, this was the case for the pressure dependence
of  the Curie temperature ($T_c$($P$)), mostly studied, up to now, in the
hole-doped part of the phase diagram of these compounds.

It was experimentally shown~\cite{Laukhin97,Moritomo97} that the pressure
coeficient (dln($T_c$)/d$P$) as a function of $T_c$ follows a sort of universal
curve, independent of the mean A ionic radius R$_A$ (1.124 \AA $<$ R$_A$ $<$
1.147 \AA)  for the A$_{0.67}$B$_{0.33}$MnO$_3$ compounds (A = Pr, Sm, Nd, Y,
La; B = Ca, Sr), or the doping level $x$ (0.2 $\leq x \leq$ 0.4) for
La$_{1-x}$Sr$_x$MnO$_3$ and Nd$_{1-x}$Sr$_x$MnO$_3$.

It has also been shown~\cite{Laukhin97} that a qualitative understanding of
this curve can be obtained in terms of the double exchange (DE)
model~\cite{Zener51,Anderson55}. Within this model, pressure increases the
transfer integral of the e$_g$ electron, which is hopping from Mn$^{3+}$ to
Mn$^{4+}$. This results in a broadening of the bandwidth (W$_0$) which yields
an increase of $T_c$. This can be understood by considering the pressure
dependence of two geometric or steric factors, both of which control W$_0$: The
Mn-O distance (d$_{Mn-O}$) and the bending angle ($\Theta$) of the Mn-O-Mn
bond. However, calculations based on a simple model related to pressure
variations of the steric factors do not give a good quantitative agreement with
the experimental data. A better quantitative prediction can be obtained by
considering the polaronic modifications of W$_0$ due to the Jahn-Teller (JT)
cooperative effect, which was clearly established for some of the manganite
members~\cite{Zhao96}. It has been shown that a positive contribution to the
pressure dependence of $T_c$ arises from the negative pressure derivative of
the isotope exponent, $\alpha = 1/2 \gamma E_{JT} / h \omega $, where $\gamma$
is a positive constant ($\leq$ 1), E$_{JT}$ is the Jahn-Teller energy and
$\omega$ an appropriate optical frequency. Even though, the experimental
results could not be suitably fitted by this model, as for example, in the
case of the studies of the oxygen-isotope effects under pressure in the
La$_{0.65}$Ca$_{0.35}$MnO$_3$ samples~\cite{Wang99}, the dlnTc/d$P$ value
observed for the $^{18}$O sample (23\%) is higher than the one predicted by
this model(6\%). It was shown~\cite{Lorenz01} that this controversy can be
solved considering the polaron theory in the intermediate electron-phonon
coupling region ($\lambda \sim$ 1) and in the adiabatic
approximation~\cite{Alex94}, which also gives an adequate description of the
electrical conductivity for temperatures above $T_c$.

The study of electron-doped manganites can shed light on the issue whether
polarons are relevant or not to the $T_c$($P$) dependence in manganites. As the
low e$_g$ level occupancy of the Mn$^{4+}$ ions no longer favors the JT
distortion, JT polaronic effects are not expected to contribute to transport
properties or to magnetic ordering in the same way they are suspected to
participate in the hole-doped part of the phase diagram. Then, a small
$T_c$($P$) dependence for the electron-doped manganites would reveal the
important role played by JT polarons in the hole-doped samples and may also
indicate, on the other hand, an active role of other type of charge carriers,
like lattice or magnetic polarons, associated with the electron-doped
manganites in other studies~\cite{Batista98,Chen01}.

For low doping levels (0.05 $\leq x \leq$ 0.15), Ca$_{1-x}$Y$_x$MnO$_3$ (CYMO)
and Ca$_{1-x}$Sm$_x$MnO$_3$ (CSMO) compounds are typical electron-doped
manganites. For this doping regime a competition between antiferromagnetic
(AF) superexchange and ferromagnetic (FM) double exchange (DE) interactions has
been revealed by dc-magnetization and electrical transport measurements as a
function of temperature and magnetic field for the CYMO system
~\cite{Garbarino01,Aliaga01,Aliaga02}. Neutron diffraction
studies~\cite{Martin99b,Martin00} for CSMO ($x$=0.10 and $x$=0.15) showed
that, besides the FM phase, a Pnma structure associated with a G-type AF is
also present in the paramagnetic state. At 10 K, the sample with $x$=0.10
exhibit a G-type AF phase, but for $x$=0.15 the C type magnetic order
(characterized by a P 2$_1$/m space group) dominates, coexisting with smaller
G AF clusters.

In this paper we study the high pressure effects on the resistivity and ac
susceptibility as a function of temperature of the low electron-doped CYMO and
CSMO manganese perovskites. Although an expected weak pressure dependence of
$T_c$($P$) was observed, a quantitative agreement could not be obtained by
solely considering the pressure sensitivity of the steric factors within a DE
scenario. Our results indicate that the characteristic hopping conduction and
the value of the pressure coefficient d(ln$T_c$)/d$P$ can be ascribed to the
polaronic nature of carriers, which can be associated with small Fr\"ohlich
polarons in a weak to intermediate coupling regime.

\section{EXPERIMENTAL}

Single phase and well oxygenated Ca$_{1-x}$Y$_x$MnO$_3$ (CYMO; $x=$0.06; 0.07;
0.08; 0.10) and Ca$_{1-x}$Sm$_x$MnO$_3$ (CSMO; $x=$0.10; 0.15) ceramic samples
were prepared by solid state reaction. This samples had been previously studied
and the details of their synthesis and characterization have been published
elsewhere~\cite{Garbarino01,Aliaga01,Maignan98,Martin99b,Mahendiran00,Respaud01,Martin99,Martin00,Vega02}.

Resistivity ($\rho$($T$)) and ac susceptibility ($\chi_{ac}$($T$)) were
measured as a function of temperature  ( 4 K $\leq T \leq$ 300 K) applying
high hydrostatic pressures up to 1 GPa. A self-clamping cell was used with a
50-50 mixture of kerosene and transformer oil as the pressure transmitting
medium. $\rho$($T$) was measured following a standard 4 terminal DC technique,
while $\chi_{ac}$($T$) was evaluated using an applied ac magnetic field of 1 Oe
at an excitation frequency of $\simeq$ 1 kHz.

\section{RESULTS AND DISCUSSION}

\begin{figure}[h]
\includegraphics[width=3in]{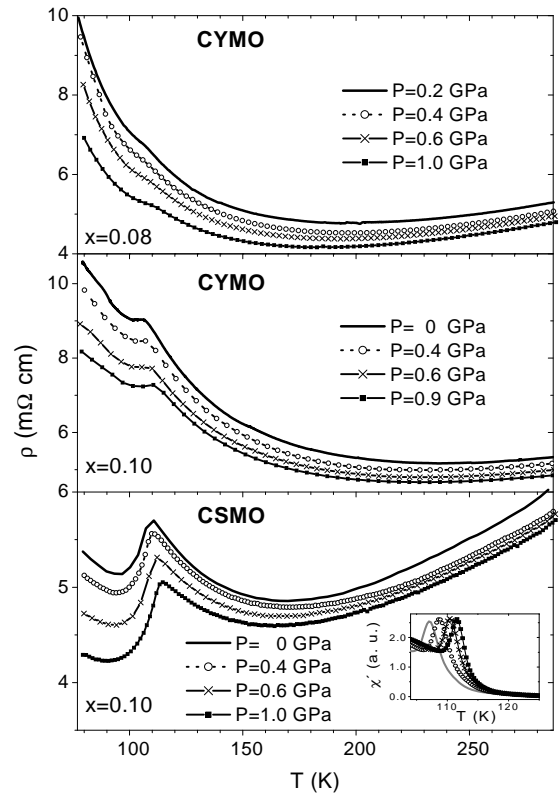}
\vspace{-5mm} \caption{Pressure dependence of the resistivity as a function of
temperature of CYMO for x=0.08 ; 0.10 and of CSMO for x=0.10. The inset shows
the peak observed in the real part of the ac susceptibility $\chi$($T$) for the
ferromagnetic transition of this latter sample. Similar transport and magnetic
results are obtained for x=0.06 (CYMO) and x=0.15 (CSMO), not shown here for
clarity.} \vspace{5mm} \label{fig:RvsTYS}
\end{figure}

Resistivity as a function of temperature for different pressures is displayed
in Fig.~\ref{fig:RvsTYS} for CYMO ($x$=0.08; 0.10) and CSMO ($x$=0.10). A
metallic conductivity at room temperature followed by a semiconducting-like
behavior when decreasing temperature can be observed for both compounds. The
low temperature resistivity shows in all cases a divergence characteristic of
an insulator behavior.

The ferromagnetic transition temperature ($T_c$) can be associated with an
inflexion or a small drop in the resistivity, depending on doping level, as
well as with a peak in the ac susceptibility, as can be noticed in the inset
of Fig.~\ref{fig:RvsTYS}. The low temperature part of  $\chi_{ac}$($T$), which
can be associated with the behavior of a cluster glass~\cite{Maignan98}, will
not be considered here, as we want to focus our study on the appearance of the
ferromagnetic ordering and its sensitivity to the applied pressure.

\begin{figure} [h]
\includegraphics[width=3.2in]{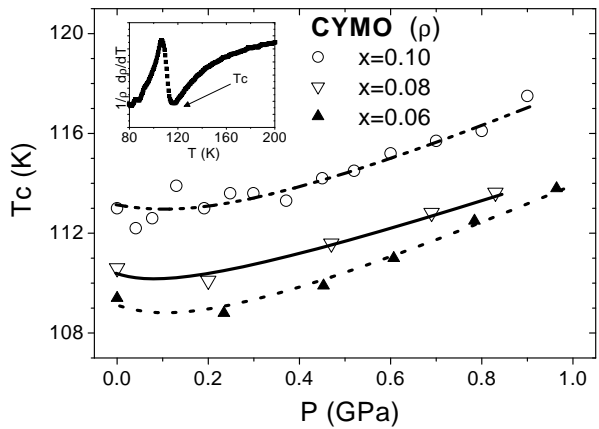}
\vspace{0mm} \caption{Pressure dependence of the ferromagnetic transition
temperature $T_c$ of CYMO. In the inset, the criteria to define $T_c$ from the
$\rho$(T) measurements is presented.} \vspace{5mm} \label{fig:TcPcymo}
\end{figure}

\begin{figure} [h]
\includegraphics[width=3.2in]{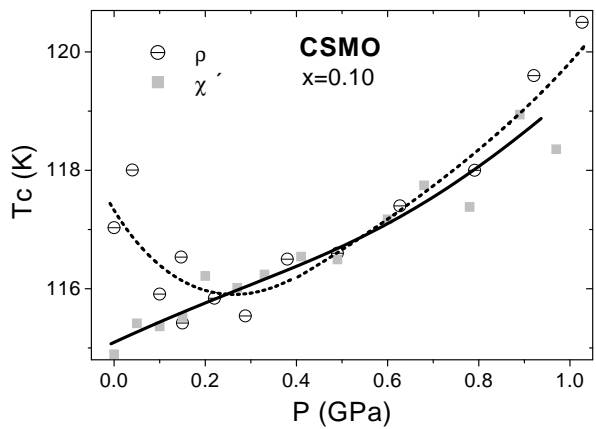}
\vspace{0mm} \caption{Comparison of the pressure dependence of the
ferromagnetic transition temperature $T_c$ of CSMO obtained using the
temperature logarithmic derivative of $\rho$(T) and $\chi$(T).} \vspace{5mm}
\label{fig:TcPcsmo}
\end{figure}

Pressure decreases resistivity and enhances the resistivity drop at $T_c$.
$T_c$ was determined using the peak of the logarithmic temperature derivative
of $\rho$($T$) and $\chi_{ac}$($T$) curves,  as illustrated in the inset of
Fig.\ref{fig:TcPcymo}. The obtained pressure sensitivity of $T_c$, depicted in
Fig.~\ref{fig:TcPcymo} and in Fig.~\ref{fig:TcPcsmo}, depends on the technique
used to characterize the sample. A non-monotonic dependence  is observed for
the resistivity data: $T_c$ remains nearly constant  for low pressures and
then softly increases for higher pressures with a slope $\sim$ 5-7
K~GPa$^{-1}$. On the other hand, susceptibility data shows essentially a
linear increase of $T_c$ with pressure with a similar slope.

\begin{figure} [h]
\includegraphics[width=3.2in]{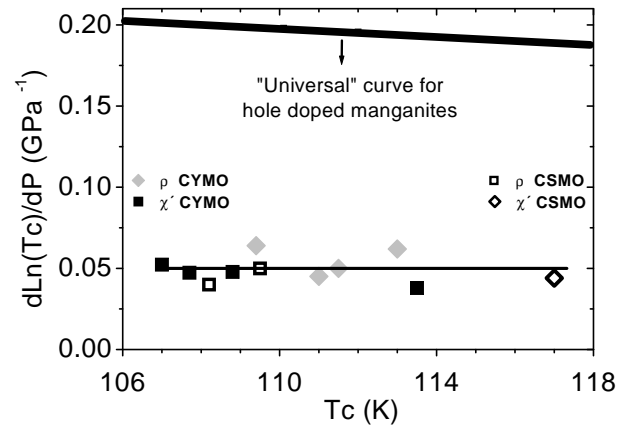}
\vspace{0mm} \caption{Dependence of the pressure coefficient dln$T_c$/d$P$ as a
function of $T_c$ for the CYMO and CSMO electron-doped manganites (determined
from $\rho$(T) or $\chi$(T) measurements).  The universal curve is a guide to
represent data of hole-doped manganites as described in the text.}
\vspace{5mm} \label{fig:dlnTc}
\end{figure}

These differences can be associated with the fact that $\rho$(T) is related
not only to the magnetic ordering of the sample but also to a percolation
problem. The percolation scenario can be generated by the coexistence of
different conducting phases and by their pressure-dependent relative
distribution. This was already demonstrated by neutron diffraction studies of
these samples, as the coexistence of FM clusters in an AFM matrix. Although the
pressure dependence of $T_c$ obtained from $\chi_{ac}$($T$) should then be more
reliable, both methods produce similar results if only the high pressure
portion of the curves is analyzed. Considering this criterion, the pressure
coefficient, dln$T_c$/d$P$, can be determined from the $T_c$($P$) curves
obtained using both techniques. As shown in Fig.\ref{fig:dlnTc}, this
coefficient is for all the samples (0.05 $\pm$ 0.01) GPa$^{-1}$, which is far
from the usual values~\cite{Moritomo97} (universal curve) obtained for
hole-doped manganites in this $T_c$ range ($\sim$ 0.20 GPa$^{-1}$).

In a DE scenario the FM transition temperature ($T_c$) is proportional to the
bare bandwidth ($W_0$), which depends on the steric parameters, d$_{Mn-O}$ and
$\Theta$($P$)~\cite{Medarde95}. It can be shown that the pressure coefficient
can then be estimated as~\cite{Laukhin97}:

\begin{equation}
\label{eq:stericPcoef} \frac{dlnT_c}{dP} = -3.5 \kappa (d_{Mn-O})-2 tan(\Theta)
\Theta \kappa (\Theta)
\end{equation}

\noindent where $\kappa (d_{Mn-O})$ and $\kappa (\Theta)$ are the bond length
and bond angle compressibilities, respectively. Using equation
\ref{eq:stericPcoef} and the values already measured for other manganese
perovskites~\cite{Radaelli97}, we can evaluate a contribution, at the most, of
0.01 GPa$^{-1}$ to the pressure coefficient produced by the variation of these
steric factors. This small value, compared with our experimental results in
Fig.\ref{fig:dlnTc}, reveals the existence of other contributions, which may
be based on the polaronic nature of electrical carriers. Though JT collective
polarons are not expected in this samples, local lattice distortions or
magnetic interactions may play an important role, which may be noticed in the
transport properties.

In this sense, the temperature dependence of the resistivity of these samples
seems to be a characteristic behavior of the electron-doped manganites, as
previously reported in the literature
~\cite{Maignan98,Neumeier00,Garbarino01,Aliaga01}. The observed dependencies
(see Fig.\ref{fig:RvsTYS}) are typical of degenerate semiconductors, where a
metallic-like conductivity can be observed at room temperature, while a
thermally activated hoppping of carriers dominates the conduction mechanism at
low temperatures. The possibility of having a Variable Range Hopping
conduction~\cite{MottDavis} was also considered, but an examination of the
data rules out this possibility.

We analyze our data considering that the electrical transport at intermediate
temperatures is effectively related to the hopping of polarons. As it was
shown for the hole-doped manganites~\cite{Lorenz01}, an adequate description
should be performed in the adiabatic limit, which must be maintained for the
electron-doped samples,  as similar characteristic times are expected to be
involved. In this case, the resistivity can be expressed as:

\begin{equation}
\label{eq:hopping} \rho(T)=\rho_0 \exp(\frac{E_a}{k_B T})
\end{equation}

\noindent with an activation energy, $E_a$, that can be approximated by

\begin{equation}
\label{eq:Ea} E_a = \frac{1}{2} E_p - J \simeq \frac{1}{2} E_p - \frac{W_0}{2z}
\end{equation}

\noindent where $E_p$ is the binding energy of the polaron, $J$ the electron
transfer integral between nearest neighbor sites,  $W_0$ is the bare
bandwidth, $z$ the lattice coordination number and $k_B$ the Boltzmann constant.

In the case that carriers could be associated with a
polaronic nature, the bare bandwidth $W_0$ should be narrowed as a consequence
of the electron-phonon interaction, producing an effective smaller $W$. $W$
and $W_0$ can be related by the following expression:

\begin{equation}
\label{eq:W} W=W_0  F^{\lambda}(E_p,\lambda,\omega_0)
\end{equation}

\noindent where $E_p$ is the polaron binding energy, $\lambda =
\frac{E_p}{W_0/2}$ the dimensionless electron-phonon coupling constant and
$\omega_0$ a characteristic optical phonon frequency. The $ F^{\lambda}$ are
different functions, defined according to the value of $\lambda$, which
determines the coupling regime~\cite{Alex94,AlexMott}. For example, in the
strong coupling regime ($\lambda \gg 1$), $F^{\lambda} \simeq \exp(-E_p/\hbar
\omega_0)$, but this function is not valid anymore for the intermediate
coupling regime as pointed out by Alexandrov and Mott~\cite{AlexMott}.

From Eqs. (\ref{eq:Ea}) and (\ref{eq:W}) we obtain that

\begin{equation}
\label{eq:EavsTc} E_a= \frac{1}{2} E_p - \frac{C T_c}{F^{\lambda}}
\end{equation}

\begin{figure} [h]
\includegraphics[width=3.2in]{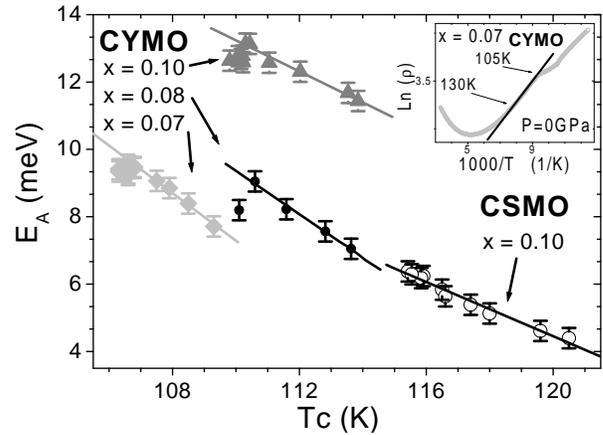}
\vspace{0mm} \caption{Activation energy $E_a$ as a function of $T_c$ for CYMO
and CSMO with different doping level. The inset shows an example of the
Ln($\rho$) vs. 1/T curves from which the $E_a$ values were extracted.}
\vspace{5mm} \label{fig:EavsTc}
\end{figure}

Following the description of Lorenz et al~\cite{Lorenz01}, we have plotted in
Fig. \ref{fig:EavsTc} the activation energy $E_a$ as a function of $T_c$
obtained for different pressures for the CYMO and CSMO with different doping
levels. Except for the lowest $T_c$ values, where, as already mentioned, their
determination is influenced by the percolation problem, an almost linear
dependence is obtained. This indicates that $E_p$ and $ F^{\lambda}$ are
essentially constants within our experimental pressure range, independently of
the pressure variation of all relevant parameters.

A linear fit of the curves of Fig. \ref{fig:EavsTc} yields an estimation of
$E_p \simeq$ 120-180 meV. The increasing value with increasing doping may point
to a tendency to localization, produced by the size-mismatch of replacing $Ca$
with small ions like $Y$ or $Sm$.  In this case, considering that $W_0 \simeq$
1 eV, the coupling constant can be estimated as $ \lambda \simeq 0.3 $, which
determines a weak to intermediate coupling regime. This value is over 4 times
smaller than the one typically obtained for the hole-doped
manganites~\cite{Dagotto01}.

Based on the fact that the electron-doped manganites are very poor conductors,
particularly for the low doping range here analyzed, we can assume the
framework introduced for doped manganites and other oxides by Alexandrov et
al~\cite{Alex99,Alex99b}, where small or lattice polarons with a long range
Fr\"ohlich interaction are the quasiparticles involved in the electrical
transport properties of these materials. Thus, for these small Fr\"ohlich
polarons we can consider the Fr\"ohlich electron-phonon coupling constant

\begin{equation}
\label{eq:alpha} \alpha= \frac{e^2}{K_p} ( \frac{m}{2 \hbar^3 \omega_0})^{1/2}
(\frac{1}{4 \pi \epsilon_0})
\end{equation}

\noindent where $ K_p^{-1} = \epsilon_{\infty}^{-1} + \epsilon_s^{-1}$ is the
effective dielectric constant and $\epsilon_s$ and $\epsilon_{\infty}$ the
static and the high frequency dielectric constants, respectively, $e$ the
electron charge,  $m$ the electron band mass and $\epsilon_0$ the vacuum
permitivity.

Following Feynman's approach for the weak to intermediate coupling
regime\cite{Feynman55,Mitra87}, we can express $E_p$ and the polaronic
effective mass, $m^*$ as a function of $\alpha$:

\begin{equation}
\label{eq:Ep} E_p = (\alpha + 0.0123 \alpha^2) \hbar \omega_0
\end{equation}

\begin{equation}
\label{eq:m*} \frac{m^*}{m} = 1 + \frac{\alpha}{6} + 0.025 \alpha^2
\end{equation}

Considering a typical phonon frequency ($\omega_0$) of the order of 400 to 500
cm$^{-1}$, estimated from Raman spectroscopy studies of CaMnO$_3$
\cite{Abrashev02}, and taking a mean value of $E_p =$ 0.15 eV from our data,
we obtain from Eqs. \ref{eq:alpha}, \ref{eq:Ep} and \ref{eq:m*} that $K_p =$ 11
$\pm$ 1 , $\alpha = $ 1.5 $\pm$ 0.1 and $m^*/m \simeq$ 1.3, respectively. The
last two values are in accordance with the assumption of being in the
intermediate coupling regime~\cite{Mitra87} (1 $\leq \alpha \leq 6$; m$^*$/m
$<$ 3). Also, assuming that the electron band mass is similar to the electron
mass ($m \sim m_e$), the polaron localization radius $R_p =$
($\hbar$/m$\omega_0$)$^{1/2}$ $\leq$ 8 $\AA \sim$ 2~d$_{Mn-O-Mn}$ determines a
polaron size in the limit of being considered small. While the radius of the
lattice distortion produced by the long range Fr\"ohlich interaction can be
estimated~\cite{MottDavis} as $R_d = 5\hbar K_p/(m e^2) \sim$ 30 $\AA$. These
values are in good accordance with the assumption of an electrical transport
mediated by small Fr\"ohlich polarons, i.e. a small size for the electron
localization and a large size for the lattice distortion. To our knowledge,
there is at the time no experimental determination of $\epsilon_{\infty}$ for
any electron-doped manganite, though a very high low-frequency dielectric
constant $\epsilon$ up to 10$^8$ was measured at room temperature in
Ca$_{1-x}$La$_x$MnO$_3$ ($x \leq$ 0.1) \cite{Zawilski02}. As $K_p \sim$ 3 was
estimated for LaMnO$_3$ \cite{Alex99}, a higher value for the electron-doped
CaMnO$_3$ can be expected, as a consequence derived from Eq. \ref{eq:alpha}
and its lower coupling constant, when compared with the former manganite. \\

At this point, we can try to make a rough approximation to determine what is
the contribution to the pressure coefficient produced by the dependence of the
bandwidth on the mass enhancement, related to the polaronic nature of
carriers. To do so, we can consider the approximation of slow polarons ($k \ll
q_p$, where $k$ and $q_p$ are the electron and the phonon momenta,
respectively)\cite{AlexMott} where the bandwidth $W$ is essentially modified as

\begin{equation}
\label{eq:Wm} W = \frac{W_0}{m^*/m}
\end{equation}

Using Eqs. \ref{eq:m*} and \ref{eq:Wm}, the pressure coefficient can then be
expressed as

\begin{equation}
\label{eq:dlnTcdP} \frac{dlnTc}{dP} \simeq  \frac{dlnW_0}{dP}  - \frac{m}{m^*}
\frac{d(m^*/m)}{d\alpha} \frac{d\alpha}{d\omega} \frac{d\omega}{dP}
\end{equation}

From Eqs.\ref{eq:alpha} and \ref{eq:Ep} and assuming that $d\omega/dP$ =
$\delta \omega_0$ (with $\delta$ $\sim$ 0.01 GPa$^{-1}$) from reference
\onlinecite{Postorino02}, we can evaluate the second term of Eq.
\ref{eq:dlnTcdP}, which gives a positive contribution of $\sim$ 0.02
GPa$^{-1}$, which is in good agreement with the results presented in
Fig.\ref{fig:dlnTc}. Therefore, these results reveal the necessity of
considering an additional contribution to the pressure coefficient besides the
one which arises out of the steric factors. The additional contribution can be
consistently related to the pressure sensitivity of intermediate-coupling
Fr\"ohlich polaronic effects on the bandwidth.

\section{CONCLUSIONS}

We have studied the pressure sensitivity of the resistivity and of the
ferromagnetic ordering of electron-doped manganese perovskites (CYMO and CSMO)
in the low doping regime. The obtained values of dln$T_c$/d$P$ are considerably
smaller than the ones measured for hole-doped manganites in the same $T_c$
range. Within a DE scenario this values can not be explained as a consequence
of the pressure variation of steric factors. A better quantitative agreement
can be obtained if a scenario of small intermediate-coupled Fr\"ohlich polarons
is considered.

\section{ACKNOWLEDGEMENTS}

We would like to acknowledge financial support by ANPCyT Grant PICT97-03-1700,
Fundaci\'on Antorchas A-13622/1-31, and UBACyT JW11. We are indebted to P.
Levy, F. Parisi and R. Zysler for fruitful discussions and to V. Bekeris and
P. Tamborenea for a critical reading of the manuscript. We also acknowledge
technical assistance from C. Chiliotte, D. Gim\'enez, E. P\'erez Wodtke and D.
Rodr\'{\i}guez Melgarejo.


\begin{thebibliography}{33}
\expandafter\ifx\csname natexlab\endcsname\relax\def\natexlab#1{#1}\fi
\expandafter\ifx\csname bibnamefont\endcsname\relax
  \def\bibnamefont#1{#1}\fi
\expandafter\ifx\csname bibfnamefont\endcsname\relax
  \def\bibfnamefont#1{#1}\fi
\expandafter\ifx\csname citenamefont\endcsname\relax
  \def\citenamefont#1{#1}\fi
\expandafter\ifx\csname url\endcsname\relax
  \def\url#1{\texttt{#1}}\fi
\expandafter\ifx\csname urlprefix\endcsname\relax\def\urlprefix{URL }\fi
\providecommand{\bibinfo}[2]{#2}
\providecommand{\eprint}[2][]{\url{#2}}

\bibitem[{\citenamefont{Laukhin et~al.}(1997)\citenamefont{Laukhin,
  Fontcuberta, Garcia-Mu\={n}oz, and Obradors}}]{Laukhin97}
\bibinfo{author}{\bibfnamefont{V.}~\bibnamefont{Laukhin}},
  \bibinfo{author}{\bibfnamefont{J.}~\bibnamefont{Fontcuberta}},
  \bibinfo{author}{\bibfnamefont{J.~L.} \bibnamefont{Garcia-Mu\={n}oz}},
  \bibnamefont{and} \bibinfo{author}{\bibfnamefont{X.}~\bibnamefont{Obradors}},
  \bibinfo{journal}{Phys. Rev. B} \textbf{\bibinfo{volume}{56}},
  \bibinfo{pages}{R10009} (\bibinfo{year}{1997}).

\bibitem[{\citenamefont{Moritomo et~al.}(1997)\citenamefont{Moritomo, Kuwahara,
  and Tokura}}]{Moritomo97}
\bibinfo{author}{\bibfnamefont{Y.}~\bibnamefont{Moritomo}},
  \bibinfo{author}{\bibfnamefont{H.}~\bibnamefont{Kuwahara}}, \bibnamefont{and}
  \bibinfo{author}{\bibfnamefont{Y.}~\bibnamefont{Tokura}},
  \bibinfo{journal}{Journal of the Physical Society of Japan}
  \textbf{\bibinfo{volume}{66}}, \bibinfo{pages}{556} (\bibinfo{year}{1997}).

\bibitem[{\citenamefont{Zener}(1951)}]{Zener51}
\bibinfo{author}{\bibfnamefont{C.}~\bibnamefont{Zener}},
  \bibinfo{journal}{Phys. Rev.} \textbf{\bibinfo{volume}{82}},
  \bibinfo{pages}{403} (\bibinfo{year}{1951}).

\bibitem[{\citenamefont{Anderson and Hasagawa}(1955)}]{Anderson55}
\bibinfo{author}{\bibfnamefont{P.~W.} \bibnamefont{Anderson}} \bibnamefont{and}
  \bibinfo{author}{\bibfnamefont{H.}~\bibnamefont{Hasagawa}},
  \bibinfo{journal}{Phys. Rev} \textbf{\bibinfo{volume}{100}},
  \bibinfo{pages}{675} (\bibinfo{year}{1955}).

\bibitem[{\citenamefont{Zhao et~al.}(1996)\citenamefont{Zhao, Conder, Keller,
  and M$\ddot{u}$ller}}]{Zhao96}
\bibinfo{author}{\bibfnamefont{G.~M.} \bibnamefont{Zhao}},
  \bibinfo{author}{\bibfnamefont{K.}~\bibnamefont{Conder}},
  \bibinfo{author}{\bibfnamefont{H.}~\bibnamefont{Keller}}, \bibnamefont{and}
  \bibinfo{author}{\bibfnamefont{K.~A.} \bibnamefont{M$\ddot{u}$ller}},
  \bibinfo{journal}{Nature} \textbf{\bibinfo{volume}{381}},
  \bibinfo{pages}{676} (\bibinfo{year}{1996}).

\bibitem[{\citenamefont{Wang et~al.}(1999)\citenamefont{Wang, Heilman, Lorenz,
  Xue, Chu, Franck, and Chen}}]{Wang99}
\bibinfo{author}{\bibfnamefont{Y.~S.} \bibnamefont{Wang}},
  \bibinfo{author}{\bibfnamefont{A.~K.} \bibnamefont{Heilman}},
  \bibinfo{author}{\bibfnamefont{B.}~\bibnamefont{Lorenz}},
  \bibinfo{author}{\bibfnamefont{Y.~Y.} \bibnamefont{Xue}},
  \bibinfo{author}{\bibfnamefont{C.~W.} \bibnamefont{Chu}},
  \bibinfo{author}{\bibfnamefont{J.~P.} \bibnamefont{Franck}},
  \bibnamefont{and} \bibinfo{author}{\bibfnamefont{W.~M.} \bibnamefont{Chen}},
  \bibinfo{journal}{Phys. Rev. B} \textbf{\bibinfo{volume}{60}},
  \bibinfo{pages}{R14998} (\bibinfo{year}{1999}).

\bibitem[{\citenamefont{Lorenz et~al.}(2001)\citenamefont{Lorenz, Heilman,
  Wang, Xue, Chu, Zhang, and Franck}}]{Lorenz01}
\bibinfo{author}{\bibfnamefont{B.}~\bibnamefont{Lorenz}},
  \bibinfo{author}{\bibfnamefont{A.~K.} \bibnamefont{Heilman}},
  \bibinfo{author}{\bibfnamefont{Y.~S.} \bibnamefont{Wang}},
  \bibinfo{author}{\bibfnamefont{Y.~Y.} \bibnamefont{Xue}},
  \bibinfo{author}{\bibfnamefont{C.~W.} \bibnamefont{Chu}},
  \bibinfo{author}{\bibfnamefont{G.}~\bibnamefont{Zhang}}, \bibnamefont{and}
  \bibinfo{author}{\bibfnamefont{J.~P.} \bibnamefont{Franck}},
  \bibinfo{journal}{Phys. Rev. B} \textbf{\bibinfo{volume}{63}},
  \bibinfo{pages}{144405} (\bibinfo{year}{2001}).

\bibitem[{\citenamefont{Alexandrov et~al.}(1994)\citenamefont{Alexandrov,
  Kabanov, and Ray}}]{Alex94}
\bibinfo{author}{\bibfnamefont{A.~S.} \bibnamefont{Alexandrov}},
  \bibinfo{author}{\bibfnamefont{V.~V.} \bibnamefont{Kabanov}},
  \bibnamefont{and} \bibinfo{author}{\bibfnamefont{D.~K.} \bibnamefont{Ray}},
  \bibinfo{journal}{Phys. Rev. B} \textbf{\bibinfo{volume}{49}},
  \bibinfo{pages}{9915} (\bibinfo{year}{1994}).

\bibitem[{\citenamefont{Batista et~al.}(1998)\citenamefont{Batista, Eroles,
  Avignon, and Alascio}}]{Batista98}
\bibinfo{author}{\bibfnamefont{C.~D.} \bibnamefont{Batista}},
  \bibinfo{author}{\bibfnamefont{J.~M.}~\bibnamefont{Eroles}},
  \bibinfo{author}{\bibfnamefont{M.}~\bibnamefont{Avignon}}, \bibnamefont{and}
  \bibinfo{author}{\bibfnamefont{B.}~\bibnamefont{Alascio}},
  \bibinfo{journal}{Phys. Rev. B} \textbf{\bibinfo{volume}{58}},
  \bibinfo{pages}{14689} (\bibinfo{year}{1998}).

\bibitem[{\citenamefont{Chen and Allen}(2001)}]{Chen01}
\bibinfo{author}{\bibfnamefont{Y.-R.} \bibnamefont{Chen}} \bibnamefont{and}
  \bibinfo{author}{\bibfnamefont{P.~B.} \bibnamefont{Allen}},
  \bibinfo{journal}{Phys. Rev. B} \textbf{\bibinfo{volume}{64}},
  \bibinfo{pages}{064401} (\bibinfo{year}{2001}).

\bibitem[{\citenamefont{Garbarino et~al.}(2001)\citenamefont{Garbarino,
  Par\'{o}n, Monteverde, Acha, Leyva, Vega, Polla, Bri\'atico, and
  Alascio}}]{Garbarino01}
\bibinfo{author}{\bibfnamefont{G.}~\bibnamefont{Garbarino}},
  \bibinfo{author}{\bibfnamefont{S.}~\bibnamefont{Par\'{o}n}},
  \bibinfo{author}{\bibfnamefont{M.}~\bibnamefont{Monteverde}},
  \bibinfo{author}{\bibfnamefont{C.}~\bibnamefont{Acha}},
  \bibinfo{author}{\bibfnamefont{G.}~\bibnamefont{Leyva}},
  \bibinfo{author}{\bibfnamefont{D.}~\bibnamefont{Vega}},
  \bibinfo{author}{\bibfnamefont{G.}~\bibnamefont{Polla}},
  \bibinfo{author}{\bibfnamefont{J.}~\bibnamefont{Bri\'atico}},
  \bibnamefont{and} \bibinfo{author}{\bibfnamefont{B.}~\bibnamefont{Alascio}},
  \bibinfo{journal}{J. Mag. Mag. Mater.} \textbf{\bibinfo{volume}{226-230}},
  \bibinfo{pages}{843} (\bibinfo{year}{2001}).

\bibitem[{\citenamefont{Aliaga et~al.}(2001)\citenamefont{Aliaga, Causa,
  Alascio, Salva, Tovar, Vega, Polla, Leyva, and Konig}}]{Aliaga01}
\bibinfo{author}{\bibfnamefont{H.}~\bibnamefont{Aliaga}},
  \bibinfo{author}{\bibfnamefont{M.~T.} \bibnamefont{Causa}},
  \bibinfo{author}{\bibfnamefont{B.}~\bibnamefont{Alascio}},
  \bibinfo{author}{\bibfnamefont{H.}~\bibnamefont{Salva}},
  \bibinfo{author}{\bibfnamefont{M.}~\bibnamefont{Tovar}},
  \bibinfo{author}{\bibfnamefont{D.}~\bibnamefont{Vega}},
  \bibinfo{author}{\bibfnamefont{G.}~\bibnamefont{Polla}},
  \bibinfo{author}{\bibfnamefont{G.}~\bibnamefont{Leyva}}, \bibnamefont{and}
  \bibinfo{author}{\bibfnamefont{P.}~\bibnamefont{Konig}}, \bibinfo{journal}{J.
  Magn. Magn. Mater.} \textbf{\bibinfo{volume}{226-230}}, \bibinfo{pages}{791}
  (\bibinfo{year}{2001}).

\bibitem[{\citenamefont{Aliaga et~al.}(2002)\citenamefont{Aliaga, Causa, Salva,
  Tovar, Butera, Alascio, Vega, Polla, Leyva, and Konig}}]{Aliaga02}
\bibinfo{author}{\bibfnamefont{H.}~\bibnamefont{Aliaga}},
  \bibinfo{author}{\bibfnamefont{M.~T.} \bibnamefont{Causa}},
  \bibinfo{author}{\bibfnamefont{H.}~\bibnamefont{Salva}},
  \bibinfo{author}{\bibfnamefont{M.}~\bibnamefont{Tovar}},
  \bibinfo{author}{\bibfnamefont{A.}~\bibnamefont{Butera}},
  \bibinfo{author}{\bibfnamefont{B.}~\bibnamefont{Alascio}},
  \bibinfo{author}{\bibfnamefont{D.}~\bibnamefont{Vega}},
  \bibinfo{author}{\bibfnamefont{G.}~\bibnamefont{Polla}},
  \bibinfo{author}{\bibfnamefont{G.}~\bibnamefont{Leyva}}, \bibnamefont{and}
  \bibinfo{author}{\bibfnamefont{P.}~\bibnamefont{Konig}}
  (\bibinfo{year}{2002}), \bibinfo{note}{submitted to J.Phys.: Condensed
  Matter}.

\bibitem[{\citenamefont{Martin et~al.}(2000)\citenamefont{Martin, Maignan,
  Hervieu, Raveau, Jir\'ak, Savosta, Kurbakov, Trounov, Andr\'e, and
  Bour\'ee}}]{Martin00}
\bibinfo{author}{\bibfnamefont{C.}~\bibnamefont{Martin}},
  \bibinfo{author}{\bibfnamefont{A.}~\bibnamefont{Maignan}},
  \bibinfo{author}{\bibfnamefont{M.}~\bibnamefont{Hervieu}},
  \bibinfo{author}{\bibfnamefont{B.}~\bibnamefont{Raveau}},
  \bibinfo{author}{\bibfnamefont{Z.}~\bibnamefont{Jir\'ak}},
  \bibinfo{author}{\bibfnamefont{M.~M.} \bibnamefont{Savosta}},
  \bibinfo{author}{\bibfnamefont{A.}~\bibnamefont{Kurbakov}},
  \bibinfo{author}{\bibfnamefont{V.}~\bibnamefont{Trounov}},
  \bibinfo{author}{\bibfnamefont{G.}~\bibnamefont{Andr\'e}}, \bibnamefont{and}
  \bibinfo{author}{\bibfnamefont{F.}~\bibnamefont{Bour\'ee}},
  \bibinfo{journal}{Phys. Rev. B} \textbf{\bibinfo{volume}{62}},
  \bibinfo{pages}{6442} (\bibinfo{year}{2000}).

\bibitem[{\citenamefont{Martin et~al.}(1999{\natexlab{a}})\citenamefont{Martin,
  Maignan, Hervieu, Raveau, Jirak, Kurbakov, Trounov, Andr\'e, and
  Bour\'ee}}]{Martin99b}
\bibinfo{author}{\bibfnamefont{C.}~\bibnamefont{Martin}},
  \bibinfo{author}{\bibfnamefont{A.}~\bibnamefont{Maignan}},
  \bibinfo{author}{\bibfnamefont{M.}~\bibnamefont{Hervieu}},
  \bibinfo{author}{\bibfnamefont{B.}~\bibnamefont{Raveau}},
  \bibinfo{author}{\bibfnamefont{Z.}~\bibnamefont{Jirak}},
  \bibinfo{author}{\bibfnamefont{A.}~\bibnamefont{Kurbakov}},
  \bibinfo{author}{\bibfnamefont{V.}~\bibnamefont{Trounov}},
  \bibinfo{author}{\bibfnamefont{G.}~\bibnamefont{Andr\'e}}, \bibnamefont{and}
  \bibinfo{author}{\bibfnamefont{F.}~\bibnamefont{Bour\'ee}},
  \bibinfo{journal}{J. Magn. Magn. Mater.} \textbf{\bibinfo{volume}{205}},
  \bibinfo{pages}{184} (\bibinfo{year}{1999}{\natexlab{a}}).

\bibitem[{\citenamefont{Mahendiran et~al.}(2000)\citenamefont{Mahendiran,
  Maignan, Martin, Hervieu, and Raveau}}]{Mahendiran00}
\bibinfo{author}{\bibfnamefont{R.}~\bibnamefont{Mahendiran}},
  \bibinfo{author}{\bibfnamefont{A.}~\bibnamefont{Maignan}},
  \bibinfo{author}{\bibfnamefont{C.}~\bibnamefont{Martin}},
  \bibinfo{author}{\bibfnamefont{M.}~\bibnamefont{Hervieu}}, \bibnamefont{and}
  \bibinfo{author}{\bibfnamefont{B.}~\bibnamefont{Raveau}},
  \bibinfo{journal}{Phys. Rev. B} \textbf{\bibinfo{volume}{62}},
  \bibinfo{pages}{11644} (\bibinfo{year}{2000}).

\bibitem[{\citenamefont{Maignan et~al.}(1998)\citenamefont{Maignan, Martin,
  Damay, Raveau, and Hejtmanek}}]{Maignan98}
\bibinfo{author}{\bibfnamefont{A.}~\bibnamefont{Maignan}},
  \bibinfo{author}{\bibfnamefont{C.}~\bibnamefont{Martin}},
  \bibinfo{author}{\bibfnamefont{F.}~\bibnamefont{Damay}},
  \bibinfo{author}{\bibfnamefont{B.}~\bibnamefont{Raveau}}, \bibnamefont{and}
  \bibinfo{author}{\bibfnamefont{J.}~\bibnamefont{Hejtmanek}},
  \bibinfo{journal}{Phys. Rev. B} \textbf{\bibinfo{volume}{58}},
  \bibinfo{pages}{2758} (\bibinfo{year}{1998}).

\bibitem[{\citenamefont{Martin et~al.}(1999{\natexlab{b}})\citenamefont{Martin,
  Maignan, Hervieu, and Raveau}}]{Martin99}
\bibinfo{author}{\bibfnamefont{C.}~\bibnamefont{Martin}},
  \bibinfo{author}{\bibfnamefont{A.}~\bibnamefont{Maignan}},
  \bibinfo{author}{\bibfnamefont{M.}~\bibnamefont{Hervieu}}, \bibnamefont{and}
  \bibinfo{author}{\bibfnamefont{B.}~\bibnamefont{Raveau}},
  \bibinfo{journal}{Phys. Rev. B} \textbf{\bibinfo{volume}{60}},
  \bibinfo{pages}{12191} (\bibinfo{year}{1999}{\natexlab{b}}).

\bibitem[{\citenamefont{Vega et~al.}(2002)\citenamefont{Vega, Ramos, Aliaga,
  Causa, B.~Alascio, Polla, Leyva, König, and Torriani}}]{Vega02}
\bibinfo{author}{\bibfnamefont{D.}~\bibnamefont{Vega}},
  \bibinfo{author}{\bibfnamefont{C.}~\bibnamefont{Ramos}},
  \bibinfo{author}{\bibfnamefont{H.}~\bibnamefont{Aliaga}},
  \bibinfo{author}{\bibfnamefont{M.~T.} \bibnamefont{Causa}},
  \bibinfo{author}{\bibfnamefont{M.~T.} \bibnamefont{B.~Alascio}},
  \bibinfo{author}{\bibfnamefont{G.}~\bibnamefont{Polla}},
  \bibinfo{author}{\bibfnamefont{G.}~\bibnamefont{Leyva}},
  \bibinfo{author}{\bibfnamefont{P.}~\bibnamefont{König}}, \bibnamefont{and}
  \bibinfo{author}{\bibfnamefont{I.}~\bibnamefont{Torriani}},
  \bibinfo{journal}{Physica B: Condensed Matter}
  \textbf{\bibinfo{volume}{320}}, \bibinfo{pages}{37} (\bibinfo{year}{2002}).

\bibitem[{\citenamefont{Respaud et~al.}(2001)\citenamefont{Respaud, Broto,
  Rakoto, Vanacken, Wagner, Martin, Maignan, and Raveau}}]{Respaud01}
\bibinfo{author}{\bibfnamefont{M.}~\bibnamefont{Respaud}},
  \bibinfo{author}{\bibfnamefont{J.~M.} \bibnamefont{Broto}},
  \bibinfo{author}{\bibfnamefont{H.}~\bibnamefont{Rakoto}},
  \bibinfo{author}{\bibfnamefont{J.}~\bibnamefont{Vanacken}},
  \bibinfo{author}{\bibfnamefont{P.}~\bibnamefont{Wagner}},
  \bibinfo{author}{\bibfnamefont{C.}~\bibnamefont{Martin}},
  \bibinfo{author}{\bibfnamefont{A.}~\bibnamefont{Maignan}}, \bibnamefont{and}
  \bibinfo{author}{\bibfnamefont{B.}~\bibnamefont{Raveau}},
  \bibinfo{journal}{Phys. Rev. B} \textbf{\bibinfo{volume}{63}},
  \bibinfo{pages}{144426} (\bibinfo{year}{2001}).

\bibitem[{\citenamefont{Medarde et~al.}(1995)\citenamefont{Medarde, Mesot,
  Lacorre, Rosenkranz, Fischer, and Gobrecht}}]{Medarde95}
\bibinfo{author}{\bibfnamefont{M.}~\bibnamefont{Medarde}},
  \bibinfo{author}{\bibfnamefont{J.}~\bibnamefont{Mesot}},
  \bibinfo{author}{\bibfnamefont{P.}~\bibnamefont{Lacorre}},
  \bibinfo{author}{\bibfnamefont{S.}~\bibnamefont{Rosenkranz}},
  \bibinfo{author}{\bibfnamefont{P.}~\bibnamefont{Fischer}}, \bibnamefont{and}
  \bibinfo{author}{\bibfnamefont{K.}~\bibnamefont{Gobrecht}},
  \bibinfo{journal}{Phys. Rev. B} \textbf{\bibinfo{volume}{52}},
  \bibinfo{pages}{9248} (\bibinfo{year}{1995}).

\bibitem[{\citenamefont{Radaelli et~al.}(1997)\citenamefont{Radaelli, Iannone,
  Marezio, Hwang, Cheong, Jorgensen, and Argyriou}}]{Radaelli97}
\bibinfo{author}{\bibfnamefont{P.~G.} \bibnamefont{Radaelli}},
  \bibinfo{author}{\bibfnamefont{G.}~\bibnamefont{Iannone}},
  \bibinfo{author}{\bibfnamefont{M.}~\bibnamefont{Marezio}},
  \bibinfo{author}{\bibfnamefont{H.~Y.} \bibnamefont{Hwang}},
  \bibinfo{author}{\bibfnamefont{S.-W.} \bibnamefont{Cheong}},
  \bibinfo{author}{\bibfnamefont{J.~D.} \bibnamefont{Jorgensen}},
  \bibnamefont{and} \bibinfo{author}{\bibfnamefont{D.~N.}
  \bibnamefont{Argyriou}}, \bibinfo{journal}{Phys. Rev. B}
  \textbf{\bibinfo{volume}{56}}, \bibinfo{pages}{8265} (\bibinfo{year}{1997}).

\bibitem[{\citenamefont{Neumeier and Cohn}(2000)}]{Neumeier00}
\bibinfo{author}{\bibfnamefont{J.~J.} \bibnamefont{Neumeier}} \bibnamefont{and}
  \bibinfo{author}{\bibfnamefont{J.~L} \bibnamefont{Cohn}},
  \bibinfo{journal}{Phys. Rev. B} \textbf{\bibinfo{volume}{61}},
  \bibinfo{pages}{14319} (\bibinfo{year}{2000}).

\bibitem[{\citenamefont{Mott and Davis}(1971)}]{MottDavis}
\bibinfo{author}{\bibfnamefont{N.~F.} \bibnamefont{Mott}} \bibnamefont{and}
  \bibinfo{author}{\bibfnamefont{E.~A.} \bibnamefont{Davis}},
  \emph{\bibinfo{title}{Electronic Processes in Non-crystalline Materials}}
  (\bibinfo{publisher}{Clarendon Press - Oxford}, \bibinfo{year}{1971}).

\bibitem[{\citenamefont{Alexandrov and Mott}(1995)}]{AlexMott}
\bibinfo{author}{\bibfnamefont{A.~S.} \bibnamefont{Alexandrov}}
  \bibnamefont{and} \bibinfo{author}{\bibfnamefont{N.~F.} \bibnamefont{Mott}},
  \emph{\bibinfo{title}{Polarons and Bipolarons}} (\bibinfo{publisher}{World
  Scientific}, \bibinfo{year}{1995}).

\bibitem[{\citenamefont{Dagotto et~al.}(2001)\citenamefont{Dagotto, Hotta, and
  Moreo}}]{Dagotto01}
\bibinfo{author}{\bibfnamefont{E.}~\bibnamefont{Dagotto}},
  \bibinfo{author}{\bibfnamefont{T.}~\bibnamefont{Hotta}}, \bibnamefont{and}
  \bibinfo{author}{\bibfnamefont{A.}~\bibnamefont{Moreo}},
  \bibinfo{journal}{Physics Reports} \textbf{\bibinfo{volume}{344}},
  \bibinfo{pages}{1} (\bibinfo{year}{2001}).

\bibitem[{\citenamefont{Alexandrov and Bratkovsky}(1999)}]{Alex99}
\bibinfo{author}{\bibfnamefont{A.~S.} \bibnamefont{Alexandrov}}
  \bibnamefont{and} \bibinfo{author}{\bibfnamefont{A.~M.}
  \bibnamefont{Bratkovsky}}, \bibinfo{journal}{J. Phys.: Condens. Matter}
  \textbf{\bibinfo{volume}{11}}, \bibinfo{pages}{L531} (\bibinfo{year}{1999}).

\bibitem[{\citenamefont{Alexandrov and Kornilovitch}(1999)}]{Alex99b}
\bibinfo{author}{\bibfnamefont{A.~S.} \bibnamefont{Alexandrov}}
  \bibnamefont{and} \bibinfo{author}{\bibfnamefont{P.~E.}
  \bibnamefont{Kornilovitch}}, \bibinfo{journal}{Phys. Rev. Lett.}
  \textbf{\bibinfo{volume}{82}}, \bibinfo{pages}{807} (\bibinfo{year}{1999}).

\bibitem[{\citenamefont{Feynman}(1955)}]{Feynman55}
\bibinfo{author}{\bibfnamefont{R.~P.} \bibnamefont{Feynman}},
  \bibinfo{journal}{Phys. Rev.} \textbf{\bibinfo{volume}{97}},
  \bibinfo{pages}{660} (\bibinfo{year}{1955}).

\bibitem[{\citenamefont{Mitra et~al.}(1987)\citenamefont{Mitra, Chatterjee, and
  Mukhopadhyay}}]{Mitra87}
\bibinfo{author}{\bibfnamefont{T.~K.} \bibnamefont{Mitra}},
  \bibinfo{author}{\bibfnamefont{A.}~\bibnamefont{Chatterjee}},
  \bibnamefont{and}
  \bibinfo{author}{\bibfnamefont{S.}~\bibnamefont{Mukhopadhyay}},
  \bibinfo{journal}{Physics Reports} \textbf{\bibinfo{volume}{153}},
  \bibinfo{pages}{91} (\bibinfo{year}{1987}).

\bibitem[{\citenamefont{Abrashev et~al.}(2002)\citenamefont{Abrashev,
  B$\ddot{a}$ckstr$\ddot{o}$m, B$\ddot{o}$rjesson, Popov, Chakalov, Kolev,
  Meng, and Iliev}}]{Abrashev02}
\bibinfo{author}{\bibfnamefont{M.~V.} \bibnamefont{Abrashev}},
  \bibinfo{author}{\bibfnamefont{J.}~\bibnamefont{B$\ddot{a}$ckstr$\ddot{o}$m}%
}, \bibinfo{author}{\bibfnamefont{L.}~\bibnamefont{B$\ddot{o}$rjesson}},
  \bibinfo{author}{\bibfnamefont{V.~N.} \bibnamefont{Popov}},
  \bibinfo{author}{\bibfnamefont{R.~A.} \bibnamefont{Chakalov}},
  \bibinfo{author}{\bibfnamefont{N.}~\bibnamefont{Kolev}},
  \bibinfo{author}{\bibfnamefont{R.~L.} \bibnamefont{Meng}}, \bibnamefont{and}
  \bibinfo{author}{\bibfnamefont{M.~N.} \bibnamefont{Iliev}},
  \bibinfo{journal}{Phys. Rev. B} \textbf{\bibinfo{volume}{65}},
  \bibinfo{pages}{184301} (\bibinfo{year}{2002}).

\bibitem[{\citenamefont{Zawilski et~al.}(2002)\citenamefont{Zawilski, Peterca,
  Cohn, and Neumeier}}]{Zawilski02}
\bibinfo{author}{\bibfnamefont{B.}~\bibnamefont{Zawilski}},
  \bibinfo{author}{\bibfnamefont{M.}~\bibnamefont{Peterca}},
  \bibinfo{author}{\bibfnamefont{J.}~\bibnamefont{Cohn}}, \bibnamefont{and}
  \bibinfo{author}{\bibfnamefont{J.}~\bibnamefont{Neumeier}},
  \bibinfo{journal}{APS meeting}  (\bibinfo{year}{2002}).

\bibitem[{\citenamefont{Postorino et~al.}(2002)\citenamefont{Postorino,
  Congeduti, Degiorgi, Iti\'e, and Munsch}}]{Postorino02}
\bibinfo{author}{\bibfnamefont{P.}~\bibnamefont{Postorino}},
  \bibinfo{author}{\bibfnamefont{A.}~\bibnamefont{Congeduti}},
  \bibinfo{author}{\bibfnamefont{E.}~\bibnamefont{Degiorgi}},
  \bibinfo{author}{\bibfnamefont{J.~P.}~\bibnamefont{Iti\'e}}, \bibnamefont{and}
  \bibinfo{author}{\bibfnamefont{P.}~\bibnamefont{Munsch}},
  \bibinfo{journal}{Phys. Rev. B} \textbf{\bibinfo{volume}{65}},
  \bibinfo{pages}{224102} (\bibinfo{year}{2002}).

\end{thebibliography}
\end{document}